\documentclass[12pt,american,english]{article}
\usepackage[T1]{fontenc}
\usepackage[latin9]{inputenc}
\usepackage[a4paper]{geometry}
\pdfoutput=1
\usepackage{amsbsy}
\usepackage{amstext}
\usepackage{graphicx}
\usepackage{setspace}
\usepackage{esint}
\makeatletter

\newcommand{\lyxmathsym}[1]{\ifmmode\begingroup\def\b@ld{bold}
  \text{\ifx\math@version\b@ld\bfseries\fi#1}\endgroup\else#1\fi}

\providecommand{\tabularnewline}{\\}

\usepackage[amssymb]{SIunits}

\@ifundefined{showcaptionsetup}{}{%
 \PassOptionsToPackage{caption=false}{subfig}}
\usepackage{subfig}
\makeatother

\usepackage{babel}
\begin{document}

\title{Particle-fluid interaction forces as the source of acceleration PDF
invariance in particle size}

\author{Yosef Meller and Alex Liberzon}

\date{\date{}}

\maketitle
\selectlanguage{american}%
\global\long\def\unt#1#2{\unit{#1}{#2}}

\selectlanguage{english}%
\begin{abstract}
The conditions allowing particle suspension in turbulent flow are
of interest in many applications, but understanding them is complicated
both by the nature of turbulence and by the interaction of flow with
particles. Observations on small particles indicate an invariance
of acceleration PDFs of small particles independent of size. We show
to be true the postulated role of particle/fluid interaction forces
in maintaining suspension. The 3D-PTV method, applied for two particle
phases (tracers and inertial particles) simultaneously, was used to
obtain velocity and acceleration data, and through the use of the
particle's equation of motion the magnitude of forces representing
either the flow or the particle interaction were derived and compared.
The invariance of PDFs is shown to extend to the component forces,
and lift forces are shown to be significant.
\end{abstract}

\section{Introduction}

Some of the most long-standing open questions in fluid dynamics pertain
to the modeling of the movement of particles in turbulence. Mixing,
deposition and resuspension of particles from a wall are examples
of processes which are topics of current research interest in a variety
of scientific fields. These and similar processes are key mechanisms
in a variety of applications. Examples include transport of lead contamination
\cite{Lankey199886}, channel dredging \cite{Peltier20111673}, indoor
distribution of allergens \cite{Kim20101854}, and design of stirred
chemical reactors \cite{Ayranci2012163}, to mention just a few.

Modeling of particle motions in turbulent flow is difficult because
it involves both the modeling of the surrounding flow field, and resulting
pressure gradients; and modeling of the particle-flow interaction,
which involves the local flow around the particle and the forces resulting
from stress applied on the particle by the flow \cite{brennen2005fundamentals}.
In the equation of motion of a particle moving in a fluid pressure
gradient appears as a term depending only on the absolute Lagrangian
acceleration of the fluid around the particle, in contrast to other
terms that depend on the \emph{relative} velocity between the particle
and the flow. In very dense suspensions of particles, the two-way
or four-way coupling which means also modeling of feedback of particles
on the flow and particle-particle interactions, are needed \cite{brennen2005fundamentals},
but these are not in the scope of the present work. 

In the research fields dealing with the suspension of particles from
the walls into a flow (hereinafter called resuspension), different
approaches were developed according to the distinction between the
importance of the pressure gradient forces (i.e. only flow) versus
the interaction forces, strongly dependent on the relative particle-fluid
velocity. For example, studies on ``dust devils'', atmospheric vortices
that entrain dust, focus on static pressure gradients as a key resuspension
mechanism (entitled ``$\Delta P$ effect''), and separately from
resuspension through aerodynamic drag effect \cite{Greeley2003,Balme2006}.
A contrary example is from the fields of aerosol and powder resuspensions
where the central approach focuses on the particle-flow interaction,
and requires detailed models of the forces applied to the surface
of the particle by the flow. Several types of models exist, some focus
particularly on particle resuspension into the flow \cite{Ziskind2006,Rabinovich2009,Reeks20011,Zhang2013103}
and some developed to model the general motion of a particle in flow
\cite{Chang2003,Rizk1985}. 

The present study focuses on the relative importance of the pressure
gradient and the particle-flow interaction forces on inertial particles
in turbulent flows. The interaction forces are several terms that
arise from the interaction of the particle with the flow due to different
mechanisms and on different time and length scales. Such terms (reviewed
in section \ref{sec:Mathematical-model}) may include the well known
Stokes form drag, the Basset history term, Faxen corrections, lift
terms (for which a possible candidate is the Saffman lift term which
arises from asymmetric stresses around the particle) and other, yet
unknown terms. Although most investigations of particle motions assume
lift terms negligible for small particles, a question whether this
result holds for turbulent flows and particles with diameter of the
order of the Kolmogorov length scale, is still a matter of scientific
debate.

Some recent numerical simulation results (e.g. \cite{duct_sim}) have
found a significant lift force, as well as recent measurements of
vorticity experienced by large particles in turbulence provide an
argument supporting the significance of lift effects \cite{large_part_rot}.
Although the models applied to the motion of small (point-like) particles
are different from those for large (e.g. compared to the Kolmogorov
length scale of turbulent flow) particles,  recent studies of particle
motion in turbulence show  that the probability density function (PDF)
of particle acceleration, when normalized to its standard deviation
(hereinafter called \emph{standardized distribution}), is independent
of the particle size. This result stands for particles of size ranging
from point-like particles of the size much smaller than the Kolmogorov
length scale to the particles of diameter several times larger than
the Kolmogorov scale. 

Calzavarini et al. \cite{calzav_accel_09} proposed that acceleration
distribution of small particles that are expected to represent the
flow field represent the acceleration (and therefore the pressure
gradient forces) experienced by the flow. However, another key mechanism
is required to explain the self-similarity of acceleration distribution
of the large particles. The authors \cite{calzav_accel_09} pointed
out that the ``drag forces'' acting on the particles require more
study in order to explain the self-similarity of standardized acceleration
PDF. The present study is presenting such explanation, linked to the
particle-fluid interaction forces, including drag and lift terms at
once. 

The fact that the normalized acceleration distribution functions are
self-similar independently of the particle size and density, at least
for the particles which are much smaller or at the order of Kolmogorov
scale, is important for modeling. The acceleration PDF obtained for
a particular turbulent flow using numerical simulations or experiments
can be translated into the distribution function for the particles,
allowing a quantitative predictions of sediment transport, dispersion
and mixing. 

This study uses a unique capability to experimentally obtain both
the turbulent velocity field in the proximity of the particles and
the trajectories of particles themselves, in order to shed some light
on the underlying distributions of the forces of inertia, pressure,
drag and lift. We show that in turbulent flow, the contribution of
lift to particle acceleration is of the same order as that of drag,
so that the common practice of neglecting lift is not applicable for
particles that are not much smaller than the Kolmogorov length scale.

Our method involves the extraction of fluid and particle velocities
and acceleration from data obtained by three-dimensional Particle
Tracking Velocimetry (3D-PTV), as described in section \ref{sec:Experimental-setup};
this data, in conjunction with the equation of motion of a particle
in flow, yields the forces from pressure gradients and from other
interactions, using a processing technique described in detail in
the following section. We follow with concluding remarks about the
relative velocity as a mechanism characterizing the flow's suspension
capacity.

\section{Materials and methods\label{sec:Mathematical-model}}

\subsection{Experimental setup\label{sec:Experimental-setup}}

The study is based on unique experimental data obtained by two three-dimensional
particle tracking velocimetry (3D-PTV) systems recording simultaneously
the turbulent flow and motion of inertial particles in the same observation
volume. The quasi-homogeneous and quasi-isotropic turbulent flow in
the observation volume is maintained by eight counter-rotating disks
on both sides of the volume. Detailed description of the experimental
set-up and the methods of analysis are given by Guala et al. \cite{guala_rot_disks}
and reproduced here for brevity.

The experiment has been carried out in a glass tank of $\unt{120\times120\times140}{\milli\metre\cubed}$
in which the flow is forced mechanically from two sides by two sets
of four rotating disks with artificial roughness elements. The observation
volume of approximately $\unt{30\times30\times30}{\milli\metre\cubed}$
was centered with respect to the forced flow domain, mid-way between
the disks. More detailed information about the flow that can be produced
by this forcing device, depicted in Figure \ref{sec:Experimental-setup},
has been presented by Liberzon et al. \cite{liberzon_rot_disks},
where smooth and baffled disks were compared.

Utilization of the 3D-PTV technique is of special importance since
the second phase (i.e. the solid particles) essentially represents
Lagrangian objects, and this experimental method allows measuring
their motion, distribution (e.g. clustering) and dynamics, along with
their interaction with the carrier fluid, measured in the same frame
of reference. 

\begin{figure}
\includegraphics[width=0.7\columnwidth]{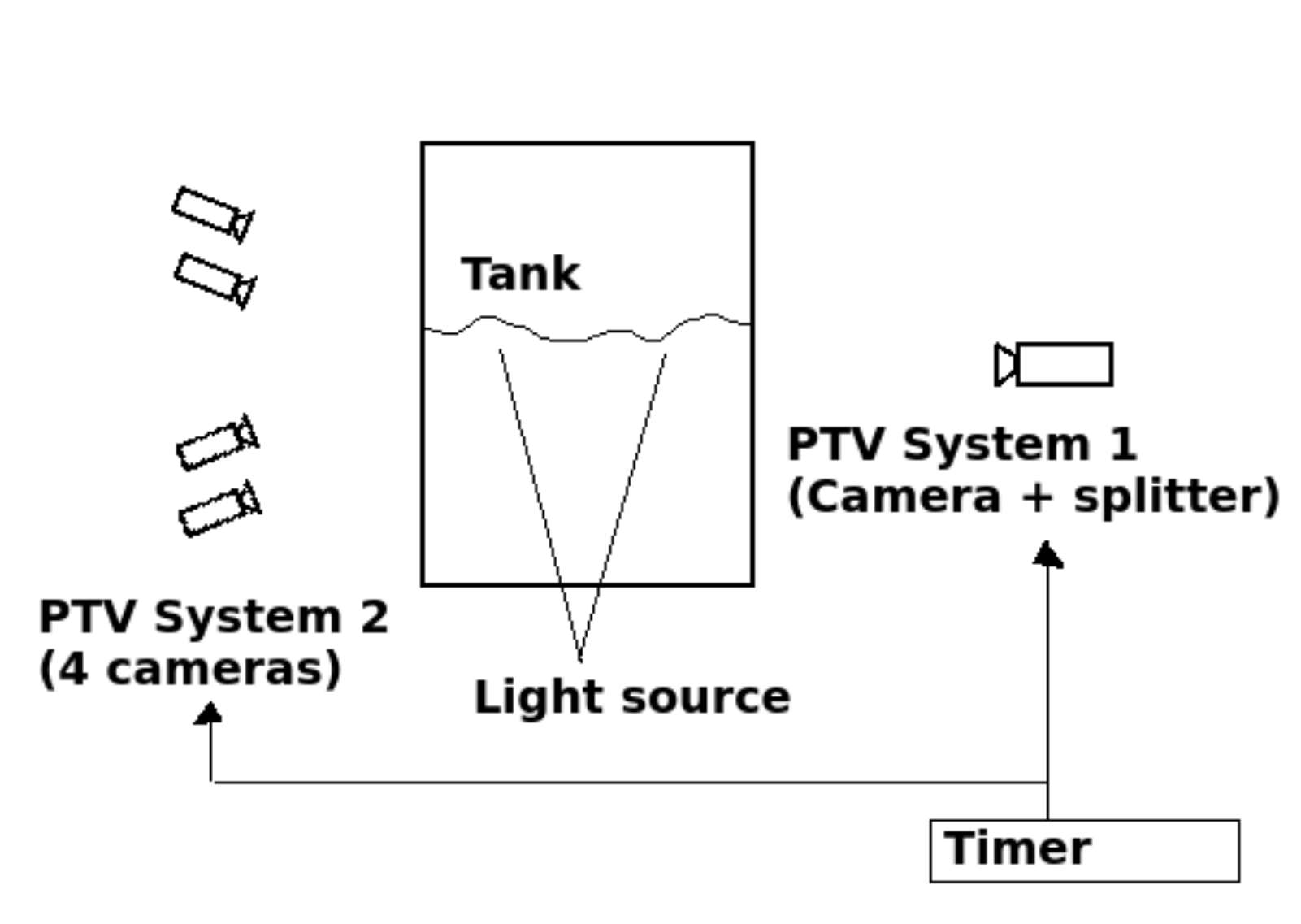} 

\caption{(Left) tank and rotating disks schematic, (Right) Experimental setup
schematic.\label{fig:Experimental-aparatus-schematic}}

\end{figure}

A sketch of the experimental apparatus is shown in Figure \ref{fig:Experimental-aparatus-schematic}
as reproduced from ref. \cite{guala_rot_disks}. Two 3D-PTV systems
were synchronized at a frame rate of 500 fps. One system, consisting
of a Photron-Ultima high-speed camera ($1024\times1024$ pixels) with
a 4-view image splitter, was set to record images of the Rhodamine
labeled silica gel particles, as the second phase. A dichroic red
filter and a low aperture were employed to filter out the light scattered
by the $\unt{40\lyxmathsym{\textendash}60}{\micro\metre}$ neutrally
buoyant polystyrene flow tracer particles (Microparticles GmbH, density
1.03 $\gram\per\centi\metre\cubed$). The second 3D-PTV system, consisting
of four MC1310 cameras (Mikrotron GmbH, $1280\times1024$ pixels),
linked to a real-time digital video recording system (IO Industries)
was used for measuring the fluid phase. These cameras were equipped
by dichroic green filters. Typical images of the tracers and the solid
particles are shown in Figure 3, where we can see that there is no
contamination by the silica particles of the images in the system
devoted to the fluid tracers, and vice versa. The observation volume
of $\unt{30\times30\times30}{\milli\metre\cubed}$ was illuminated
by a continuous $\unt{20}{\watt}$ Ar-Ion laser (Spectra-Physics).

Ref. \cite{guala_rot_disks} had measured different combinations of
tracers and inertial particles, from which we present results of one
data set using spherical porous glass particles with mean diameter
of 500 micrometers and density of 1.45 $\gram\per\centi\metre\cubed$.
For the details of the 3D-PTV method and data processing the readers
are referred to Refs. \cite{guala_rot_disks,beat_3dptv}. 

Raw data obtained by 3D-PTV may contain some measurement noise. Similarly
to the previous works (e.g. \cite{beat_3dptv,voth_variance}) the
position noise is filtered using a Savitsky-Golay filter \cite{numerical_recipes}
applied to each Cartesian component of the position along particle
trajectories.

It has been shown \cite{voth_variance} that a fit time of the order
of the Kolmogorov time scale is sufficient to filter out noise without
losing data. Therefore it is sufficient in our case to use a cubic
polynomial over a 5-point time window in the Savitsky-Golay filter.

\subsection{Equation of motion of a particle}

A particle \emph{equation of motion}, i.e. an equation that balances
the acceleration of the particle times its mass, $m_{p}\mathbf{{a}}_{p}$
with the sum of forces acting on it, can be obtained from first principles
by integrating the Navier-Stokes equation over a control volume around
a particle. 

The Navier-Stokes equations are notoriously nonlinear and hard to
handle in all but a small class of simple cases. When applied to turbulent
flow, they require a combination of assisting mathematical concepts,
and a judicious use of scale relations. It is therefore inherent to
the source of this equation, that several derivations are possible,
depending on what can be assumed about the problem at hand and which
terms can be neglected. In many derivations, the common concept that
arises from the mathematical development is that of \emph{undisturbed}
fluid velocity, i.e. the velocity that the fluid would have had at
the absence of the particle. In this way the flow is separated into
the flow field as it would have been without particles, and the disturbance
field.

This has allowed for an important application of the particle equation
of motion: reconstruction of the undisturbed flow field from measurement
of tracer particles \cite{Chang2003}. The disturbance field created
by the tracers is considered negligible compared to the undisturbed
field, in a quantifiable way \cite{mei_tracer_fidelity,tracers_melling}.

Another application of the particle motion equation is for incorporating
particle motion in numerical simulations (e.g. \cite{mattson:045107}
and many others). When simulating particles in flow, a common approach
is to separate the modeling of the undisturbed flow field from the
modeling of particle motion, and use coupling of Eulerian flow simulation
with Lagrangian particle tracking, whose underlying equation is the
particle equation of motion.

The often-quoted review work of Maxey and Riley \cite{eq_motion_maxey_riley_83}
provided two derivations: the state-of-the-art one of Corrsin and
Lumley \cite{corrsin_lumley}, and the corrected original derivation.
The Corrsin-Lumley form is
\begin{eqnarray}
m_{\mathrm{p}}\frac{dV_{i}}{dt} & = & m_{\mathrm{f}}\left(\frac{Du_{i}}{Dt}-\nu\nabla^{2}u_{i}\right)-\frac{1}{2}m_{\mathrm{f}}\frac{d}{dt}\left(V_{i}-u_{i}\right)-6\pi a\mu\left(V_{i}-u_{i}\right)\label{eq:corrsin-lumley}\\
 &  & -6\pi a^{2}\mu\int_{-\infty}^{t}\frac{\frac{d}{d\tau}\left(V_{i}-u_{i}\right)}{\sqrt{\pi\nu\left(t-\tau\right)}}d\tau+g_{i}\left(m_{\mathrm{p}}-m_{\mathrm{f}}\right);\qquad i=1,2,3\nonumber 
\end{eqnarray}
using the following variables: for the particle, its velocity $V_{i}$,
radius $a$ and particle mass $m_{\mathrm{p}}$; for the carrier fluid,
the undisturbed fluid velocity at the particle centre $u_{i}$, the
mass of fluid displaced by the particle, $m_{\mathrm{f}}$, the fluid
viscosity $\mu$ and its kinematic viscosity $\nu$.

An important notation convention is that the derivatives $d/dt$ and
$D/Dt$ represent Lagrangian derivatives, following the particle and
the containing fluid element respectively, so that (bold face symbols
denote the vector quantities): 
\begin{eqnarray}
\frac{d}{dt} & = & \frac{\partial}{\partial t}+\mathbf{V}\cdot\nabla\\
\frac{D}{Dt} & = & \frac{\partial}{\partial t}+\mathbf{u}\cdot\nabla
\end{eqnarray}

The terms on the right side of Eq. \ref{eq:corrsin-lumley} are the
different force terms acting on the particle: the pressure-gradient
force $F_{\mathrm{p}}$, added-mass force $F_{\mathrm{a}}$, Stokes
drag $F_{\mathrm{S}}$, Basset history term $F_{\mathrm{B}}$, and
buoyancy $F_{\mathrm{g}}$, respectively.

Maxey and Riley \cite{eq_motion_maxey_riley_83} kept the buoyancy
term, but replaced the other terms with expressions different by the
undisturbed velocity Laplacian (also known as the \emph{Faxen correction}s).
Recent simulations have shown \cite{Calzavarini2012237} that the
Faxen correction becomes significant only for particles with diameter
of several times the dissipative (Kolmogorov) length scale of the
flow; in our experiments, the inertial particle size is of the order
of the dissipative length scale. Therefore we use here the form of
the terms that does not contain the Faxen corrections, implying that
the pressure term can be estimated from the Lagrangian acceleration
of the fluid:
\begin{eqnarray}
\mathbf{F}_{\mathrm{p}}^{\prime} & \equiv & m_{\mathrm{f}}\frac{D\mathbf{u}}{Dt}\label{eq:pressure_force}
\end{eqnarray}

An important term that is not included in these equations of motion
which were also derived in parallel by Gatignol \cite{eq_motion_gatignol}
for the point-like particles   is the lift force. This term can, for
instance, be formulated as shear-induced lift, suggested by Saffman
\cite{lift_saffman_65}. If we add the definition of the relative
velocity $W_{i}=V_{i}-u_{i}$ of the particle, then the lift force
is proportional to the cross-product of the relative rotation with
the relative velocity: 
\begin{equation}
\mathbf{F}_{\mathrm{Sa}}\propto(\nabla\mathbf{W})\times\mathbf{W}
\end{equation}
This is not the only possible formulation of the lift force(s) that
can arise from asymmetric stresses on the surface of the particle,
and the asymmetry is expressed using the relative velocity gradient
or alternatively using the local undisturbed flow vorticity. The Saffman
lift term has been found negligible in laminar flows but may be important
in turbulent flows, where fluctuating velocity and vorticity are likely
to induce asymmetric stresses.

For the purpose of the present study, the pressure-gradient force
on the right-hand side and the total force on the particle on the
left-hand side are the only terms that are of interest individually,
hence the terms relating to particle-fluid interaction will be grouped
together into a so-called \emph{interaction force} term $\mathbf{F}_{r}\equiv\mathbf{F}_{\mathrm{S}}+\mathbf{F}_{\mathrm{B}}+\mathbf{F}_{\mathrm{Sa}}$.
Hence the equation of motion may be rewritten in the short form used
here throughout,
\begin{equation}
m_{\mathrm{p}}\frac{d\mathbf{\mathbf{V}}}{dt}-\mathbf{F}_{p}^{\prime}-\mathbf{F}_{a}-\mathbf{F}_{g}-\mathbf{F}_{r}=0
\end{equation}

A similar method of analysis has been employed previously by Sridhar
and Katz \cite{bubbles_sridhar_katz} in their study of bubble motion
in a laminar vortex. The authors \cite{bubbles_sridhar_katz} have
decomposed $\mathbf{F}_{r}$ into two orthogonal components: a the
drag force, parallel to the relative velocity, $\mathbf{F}_{\mathrm{D}}\parallel\mathbf{v}$,
and the lift force perpendicular to the relative velocity, $\mathbf{F}_{\mathrm{L}}\perp\mathbf{w}$.
The decomposed terms provided the drag and lift coefficients that
were expressed as functions of the local particle Reynolds number. 

Although the method of Ref. \cite{bubbles_sridhar_katz} appears promising
and it has been expanded from spherical bubbles to ellipsoidal ones
\cite{ford:178}, it has not been further developed for turbulent
flows.  The original work of Ref. \cite{bubbles_sridhar_katz} uses
constructions suitable for laminar flow, and makes use of an interpolation
method to obtain undisturbed velocity without examining its accuracy.
The merit of the method was demonstrated using a comparison of measured
trajectories to trajectories predicted from theory, but this approach
is not feasible in turbulent flows. For this reason we will present
a different method to compare interpolation methods, explained in
section \ref{sub:Undisturbed-fluid-velocity}. 

An additional simplification introduced in the work of \cite{bubbles_sridhar_katz}
was to include the added mass force term in the left hand side term
and create the so-called \emph{inertia force}, 
\begin{equation}
\mathbf{F}_{\mathrm{I}}=m_{\mathrm{p}}\frac{d\mathbf{v}}{dt}-\mathbf{F}_{a}
\end{equation}
For the purpose of this work we can also treat the particle and added
mass together as a particle of increased mass which is exposed to
the same pressure gradient, and the particle-fluid interaction forces
are derived from the velocity gradients around the particle. Following
this notation, the form of the equation of motion used henceforth
is the following:
\begin{equation}
\mathbf{F}_{\mathrm{I}}-\mathbf{F}_{p}^{\prime}-\mathbf{F}_{g}-\mathbf{F}_{r}=0\label{eq:final}
\end{equation}
and the interaction force is decomposed into two orthogonal components,
$\mathbf{F}_{r}=\mathbf{F}_{\mathrm{D}}+\mathbf{F}_{\mathrm{L}}$

\subsection{Undisturbed fluid velocity interpolation\label{sub:Undisturbed-fluid-velocity}}

The undisturbed fluid velocity field $\boldsymbol{u}$ is a central
concept in all the aforementioned derivations. It is assumed that
the flow tracers (i.e. small, neutrally buoyant and inertialess particles)
velocity $\mathbf{V}$ represents the fluid flow velocity at their
locations. However, in most two phase flows, the second phase (bubbles,
droplets, inertial particles, etc.) cannot be approximated as flow
tracers and the undisturbed fluid flow velocity at their position
must be obtained through interpolation of the fluid velocity of the
flow tracers at positions close to the particle.

In Ref. \cite{bubbles_sridhar_katz} neighboring bubbles were used
to estimate the fluid velocity in the vicinity of any other bubble,
serving both as flow tracers and as a second phase. The measurements
were performed in two-dimensional settings, and a convenient bilinear
interpolation method was used to obtain the undisturbed fluid velocity
field. For a three dimensional case, a different interpolation method
is needed. Moreover, we need to evaluate possible sources of error
-- if the ``flow tracers'' (for example, the small bubbles) used
in Ref. \cite{bubbles_sridhar_katz}, are not ideal tracers in all
flow conditions (e.g., in high strain regions), then the undisturbed
flow velocity at the position of the particle from the second phase,
$u_{i}(X(t),t)$ would contain both the interpolation error and the
error due to the imperfection of the ``flow tracers''. In order
to develop the method of Sridhar and Katz \cite{bubbles_sridhar_katz}
for turbulent flows, we performed a careful assessment of the three
dimensional interpolation technique.

There are many interpolation methods that could conceivably be used.
Some interpolation methods are generally applicable to any interpolation
problem, and some are tailored for preserving certain properties of
the flow (e.g. limited acceleration variance, or zero velocity divergence);
the later, however, often come at the expense of other properties
of the flow. A comprehensive survey of the suitability of all interpolation
methods to our task is a very broad undertaking. Out of all the methods
we tested, we chose the \emph{Inverse Distance Weighting} interpolation
\cite{Shepard:1968:TIF:800186.810616}, which is computationally simple
and achieves good results compared to other methods. The interpolated
value under this method is 
\[
\boldsymbol{u}_{i}=\frac{\sum_{j}u_{j}r_{ij}^{-p}}{\sum_{j}r_{ij}^{-p}}
\]
where the exponent $p$ is an arbitrary free parameter. In this study
we are able to do a careful analysis by comparing tracers velocity
to interpolated values using the same set of flow tracers. Since flow
tracers presumably have the same velocity as that of the carrying
fluid, the optimal power $p$ would minimize the least-squares difference
between the particle velocity and the interpolated velocity at the
same position. Ideally, in error-free measurements one could obtain
null difference for ideal tracers, or estimate the correct relative
velocity for real tracer particles. For a full set of $n$ flow tracers
velocity measurements for all the frames, we minimize
\begin{equation}
S=\frac{1}{n}\sum_{j=1}^{n}\sqrt{W_{i}\left(j\right)W_{i}\left(j\right)}\label{eq:self_test}
\end{equation}
 where $W_{i}\left(j\right)$ is the fluid-relative velocity vector
for particle $j$. The value minimizing $S$ in our data is $p=1.5$.

\section{Results and discussion\label{sec:Results-and-discussion}}

\subsection{Relative velocity\label{sub:Relative-velocity}}

\begin{figure}
\subfloat[Probability distribution\label{fig:Probability-distribution-rel-vel}]{\includegraphics[width=0.49\columnwidth]{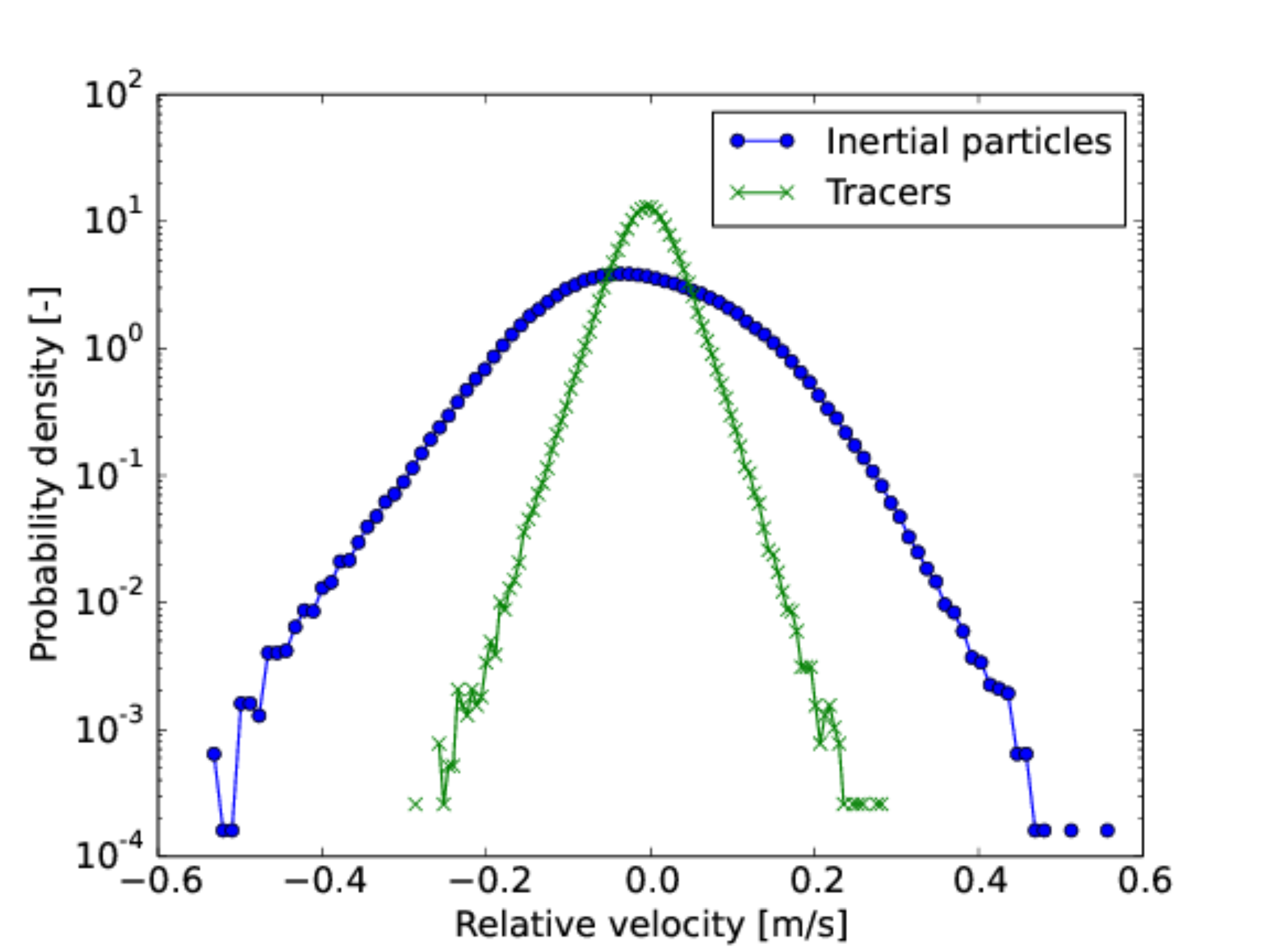}

} \subfloat[Standardized probability distribution\label{fig:Normalized-probability-distribut-rel-vel}]{\includegraphics[width=0.49\columnwidth]{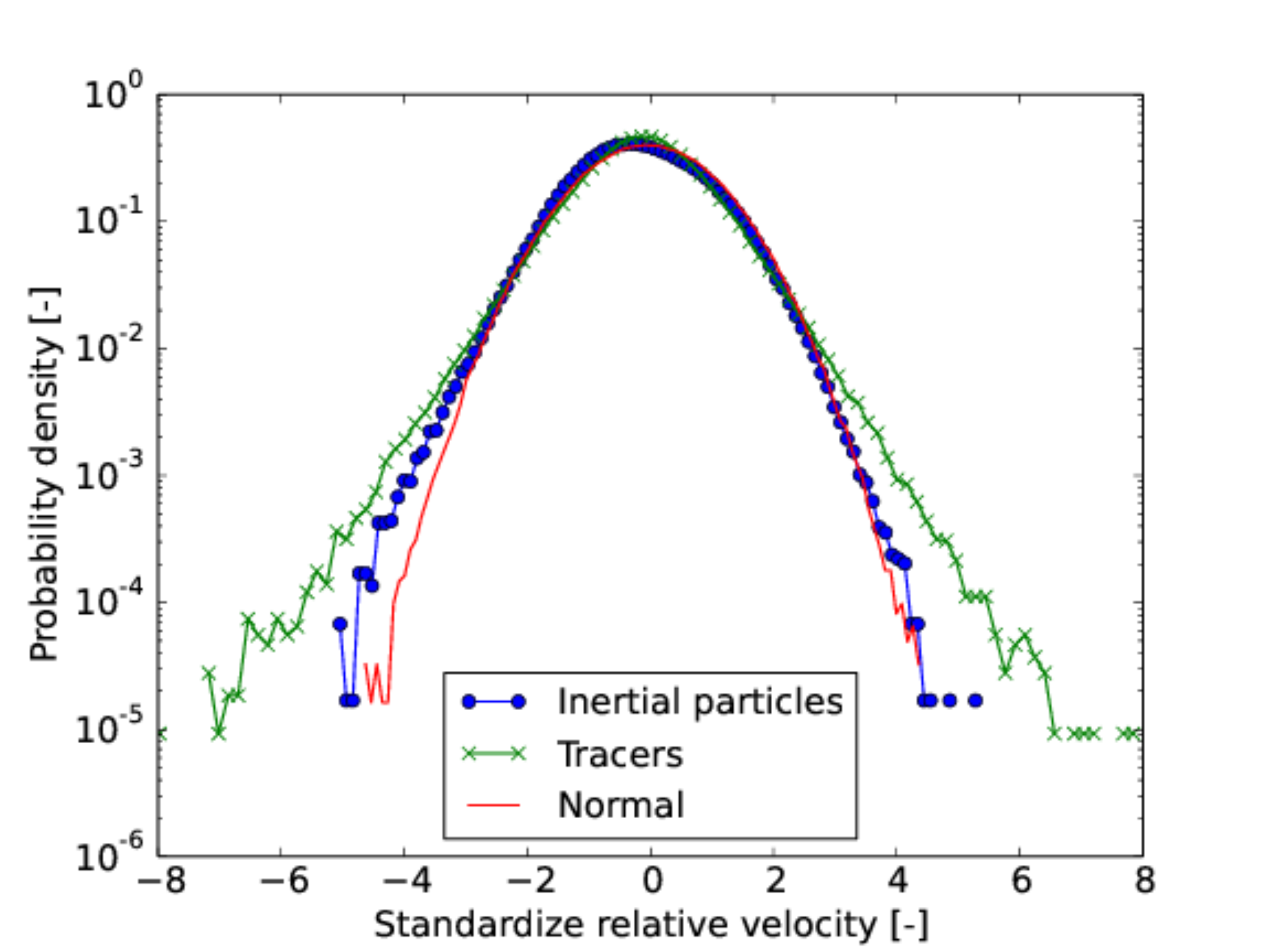}

}

\caption{Relative velocity of tracers and inertial particles.}

\end{figure}

Figure \ref{fig:Probability-distribution-rel-vel} shows the relative
velocity between particles and the fluid surrounding them. For the
tracers, the same set of particles is used both for the fluid velocity
interpolation and for the tracer velocity measurement. Although the
tracer self-test, given in Eq. \ref{eq:self_test}, minimized the
average relative velocity to $\unt{0.07}{\metre\per\second}$, the
distribution of tracers relative velocity has tails going out to $\unt{\pm0.2}{\metre\per\second}$,
for very few events. This represents the addition of tracer nonideality
to the interpolation error. The tracers are found throughout the flow
and encounter different flow conditions, some of them expose a relatively
slow dynamic response even for the tracers. The inertial particles
have a wider relative velocity distribution, showing that the heavier
particles are also slower to respond to the flow, as one might expect.

In figure \ref{fig:Normalized-probability-distribut-rel-vel}, the
velocity PDFs are normalized by the standard deviation of each distribution,
providing a normalized, or \emph{standardized}, distribution. Although
the PDFs are then close, they do not exactly ``collapse'' as is
the case of the force PDFs (section \ref{sub:Forces-on-the-particles}).
Especially in the tails of the distribution, tracers' normalized relative
velocities are somewhat higher. It can be seen from the differences
between the two shapes of the standardized distributions that flow
tracers experience some peculiar flow conditions or flow regions that
inertial particles avoid or filter out.

The fact that the tracers have a non-negligible relative velocity,
although significantly smaller than that of inertial particles, indicates
that there would be some error in undisturbed velocity estimate that
stems from the possibility of tracers not following the flow faithfully.
Apparently these extreme cases (turbulent events that correspond to
the tails of the distribution) will contribute to the noise in the
PDFs of force terms acting on inertial particles, as shown in section
\ref{sub:Forces-on-the-particles}. However, only a negligibly small
fraction of samples shows a significant relative velocity, and therefore
the noise appears only toward the tails of the force distributions
and does not affect the outcome of the study.

\begin{figure}
\noindent \begin{centering}
\includegraphics[width=0.5\columnwidth]{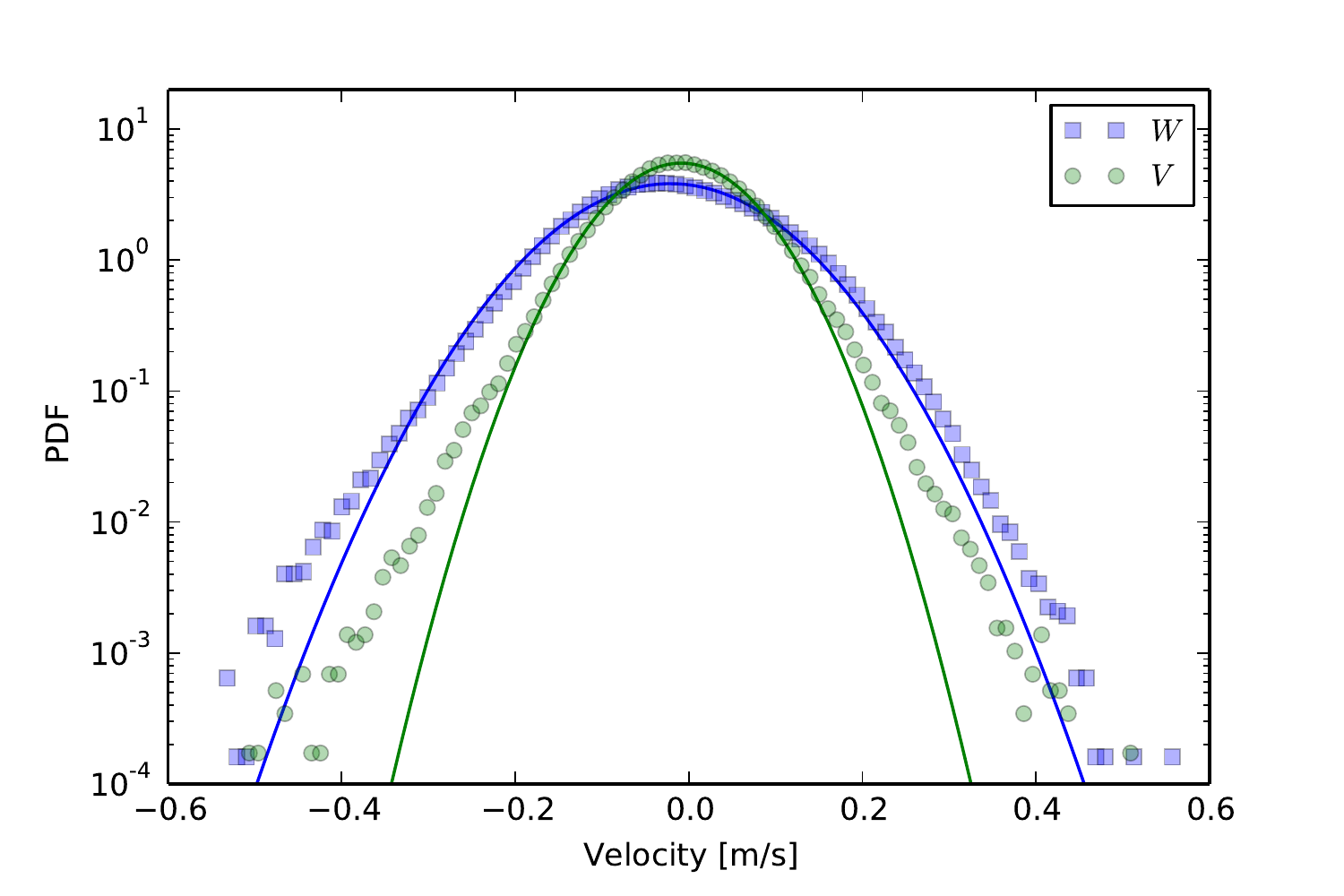}\includegraphics[width=0.5\columnwidth]{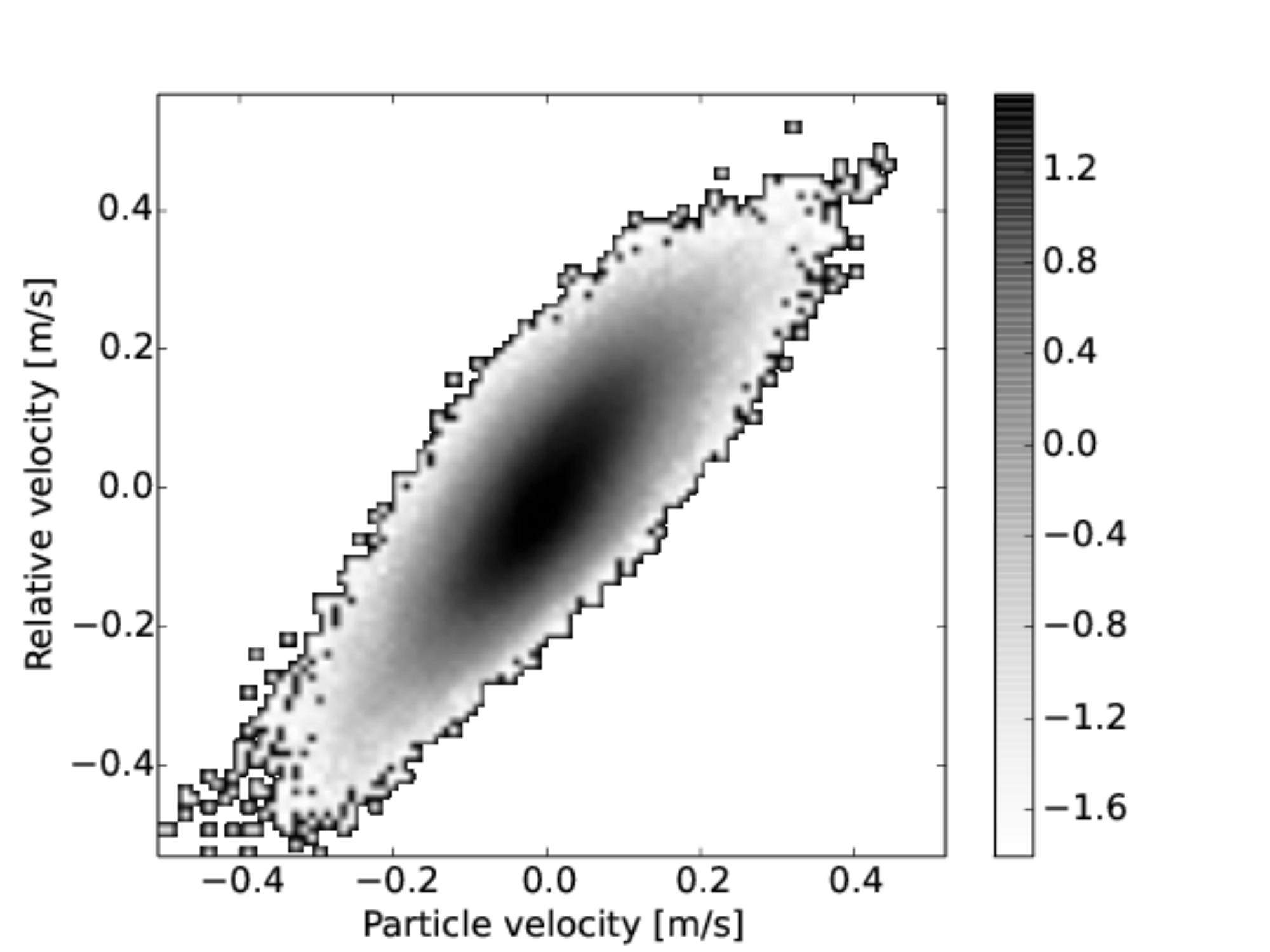}
\par\end{centering}

\caption{(Left) Absolute vs. relative velocity for inertial particles. (Right)
Joint PDF of particle velocity and relative velocity (probability
density is log scaled).\label{fig:Absolute-vs.-relative}}
\end{figure}

In addition, our experimental data of synchronously measured flow
velocity and the motion of the inertial particles, provides the option
to directly probe the commonly applied assumption on the relative
velocity of the inertial particles, $W$, to be of the same order
of magnitude as their absolute Lagrangian velocity, $V$, e.g. Ref.
\cite{large_part_rot}. The PDFs of relative and absolute velocities
of inertial particles are shown in figure \ref{fig:Absolute-vs.-relative},
together with the Gaussian fit for each case. The standard deviation
of the relative velocity is $\sigma_{\mathrm{W}}=\unt{0.105}{\metre\per\second}$
and it is slightly higher than that of the absolute velocity, $\sigma_{\mathrm{V}}=\unt{0.076}{\metre\per\second}$.
The velocities are of the same order of magnitude, yet the distributions
show that the relative velocity has broader tails, which is a sign
of intermittent turbulent flow in its proximity. The differences between
relative and absolute velocity of inertial particles is a sort of
measure of its low-pass filtering effect. Except these strong turbulence
regions (or time events)  the relative velocity and the absolute velocity
of the inertial particles are correlated statistically. Figure \ref{fig:Absolute-vs.-relative}
shows this in a joint PDF of the velocity and the relative velocity.
The correlation is close to 1, although not perfect as we can find
events with low absolute and high relative velocity, and vice versa. 

\subsection{Forces on the particles in turbulent flow\label{sub:Forces-on-the-particles}}

The major focus of this study is on the estimates of the force terms
acting on inertial particles along their trajectories in a turbulent
flow. The key method is the analysis of the forces acting on inertial
particles as compared to those acting on flow tracers, while both
estimates obtained using the same dataset and the same interpolation.
Comparing forces acting on tracers and inertial particles effectively,
we need to compensate for the different size and density of tracers
as compared to inertial particles. Since the absolute values are of
different orders of magnitude, we will normalize each probability
distribution function (PDF) to its standard deviation, emphasizing
the shape of the distribution. 

The standard deviation values used to normalize the PDFs in this section
are noted in table \ref{tab:Standard-deviations}. The pressure force
scaling ratio is of interest in that it serves as an indication of
the way tracers see the flow as opposed to inertial particles. Since
the pressure force is defined as $m_{\mathrm{f}}\frac{D\mathbf{u}}{Dt}$
(eq. \ref{eq:pressure_force}), it should scale by the mass ratio
if the distribution of $\frac{D\mathbf{u}}{Dt}$ seen by tracers is
the same as that seen by inertial particles. Our measurement indicates
that the scaling is close to the mass ratio but somewhat smaller.
Some of of the difference may be a result of a certain size distribution
of the particles around the nominal value. The rest of the difference
is linked with the low-pass filtering effect discussed in section
\ref{sub:Relative-velocity}. 

The scaling of inertia force, conversely, is larger than the mass
ratio. This shows that as a particle grows larger, the role of surface
forces grows relative to the role of the pressure force. The particle
follows the flow less closely, and hence is affected more by forces
related to the relative velocity and its gradient, i.e. drag and lift.
This is also seen by the scaling of the drag and lift forces, which
are also larger than the mass ratio.

\begin{table}
\begin{tabular}{|c|c|c|c|c|}
\hline 
Force & Inertial particles & Tracers & Inertial/Tracers ratio & Figure\tabularnewline
\hline 
\hline 
Inertia ($F_{\mathrm{I}}$) & $\unt{1.08\cdot10^{-6}}{\newton}$ & $\unt{6.55\cdot10^{-10}}{\newton}$ & $0.165\cdot10^{4}$ & \ref{fig:Probability-distributioninertia}\tabularnewline
\hline 
Pressure ($F_{\mathrm{p}}^{\prime}$) & $\unt{0.31\cdot10^{-6}}{\newton}$ & $\unt{2.38\cdot10^{-10}}{\newton}$ & $0.130\cdot10^{4}$ & \ref{fig:Normalized-probability-distribut-pressure}\tabularnewline
\hline 
Total interaction ($F_{\mathrm{r}}$) & $\unt{1.08\cdot10^{-6}}{\newton}$ & $\unt{6.55\cdot10^{-10}}{\newton}$ & $0.162\cdot10^{4}$ & \ref{fig:Standardized-interaction-force}\tabularnewline
\hline 
Drag ($F_{\mathrm{D}})$ & $\unt{0.78\cdot10^{-6}}{\newton}$ & $\unt{4.46\cdot10^{-10}}{\newton}$ & $0.175\cdot10^{4}$ & \ref{fig:Probability-distribution-drag}\tabularnewline
\hline 
Lift ($F_{\mathrm{L}}$) & $\unt{0.84\cdot10^{-6}}{\newton}$ & $\unt{5.28\cdot10^{-10}}{\newton}$ & $0.159\cdot10^{4}$ & \ref{fig:Normalized-probability-distribut-lift}\tabularnewline
\hline 
\end{tabular}

\caption{Standard deviations of the forces defined in eq. \ref{eq:final}.
The mass ratio is $0.14\cdot10^{4}$. Buoyancy force is $\unt{2.79\cdot10^{-7}}{\newton}$
for the inertial particles and $\unt{1.96\cdot10^{-11}}{\newton}$
for the tracers, both in the negative vertical direction. \label{tab:Standard-deviations}}

\end{table}

\begin{figure}
\subfloat[Standardized inertia force PDFs for inertial particles and tracers.\label{fig:Probability-distributioninertia}]{\includegraphics[width=0.49\columnwidth]{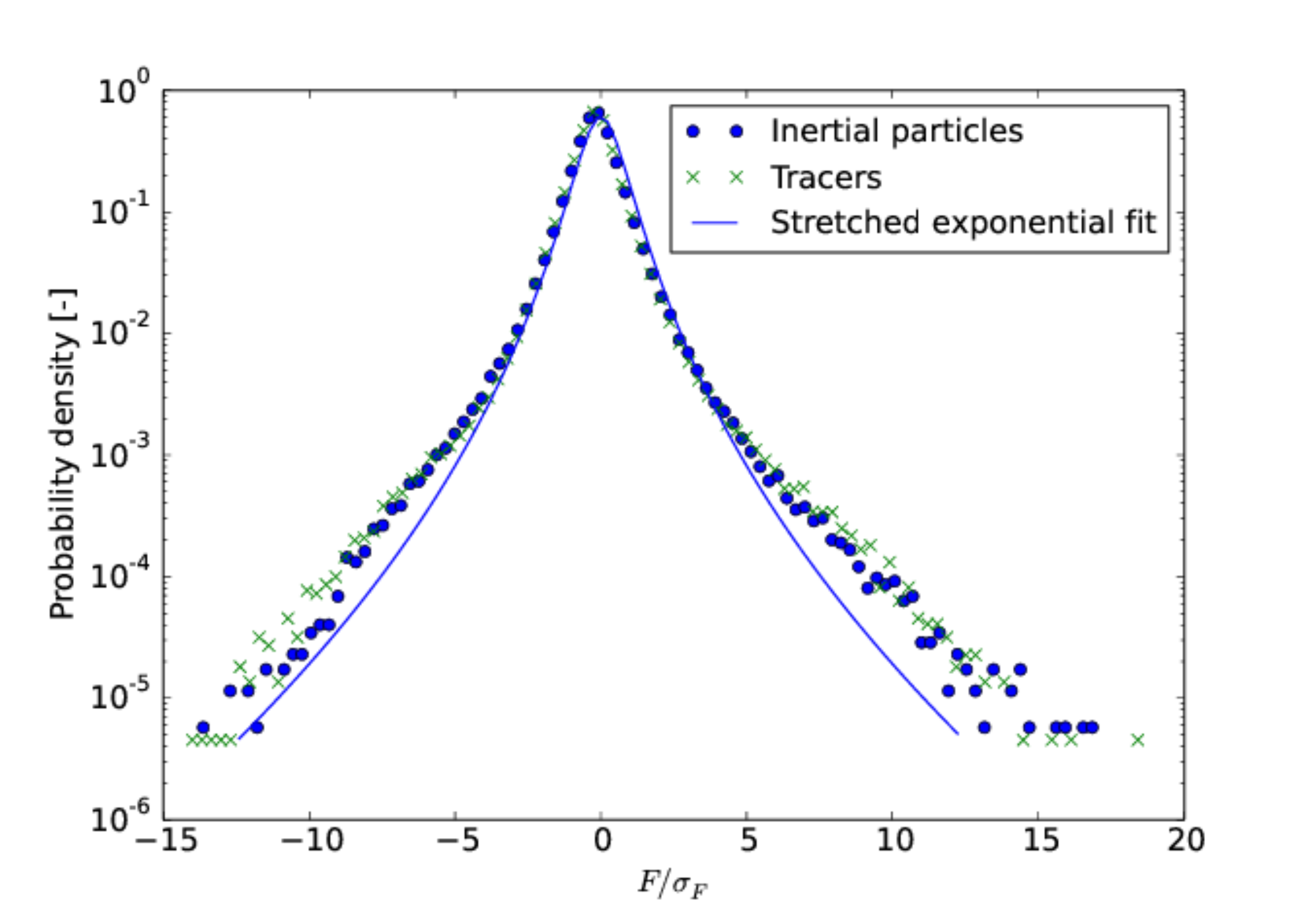}

} \subfloat[PDFs of pressure force.\label{fig:Normalized-probability-distribut-pressure}]{\includegraphics[width=0.49\columnwidth]{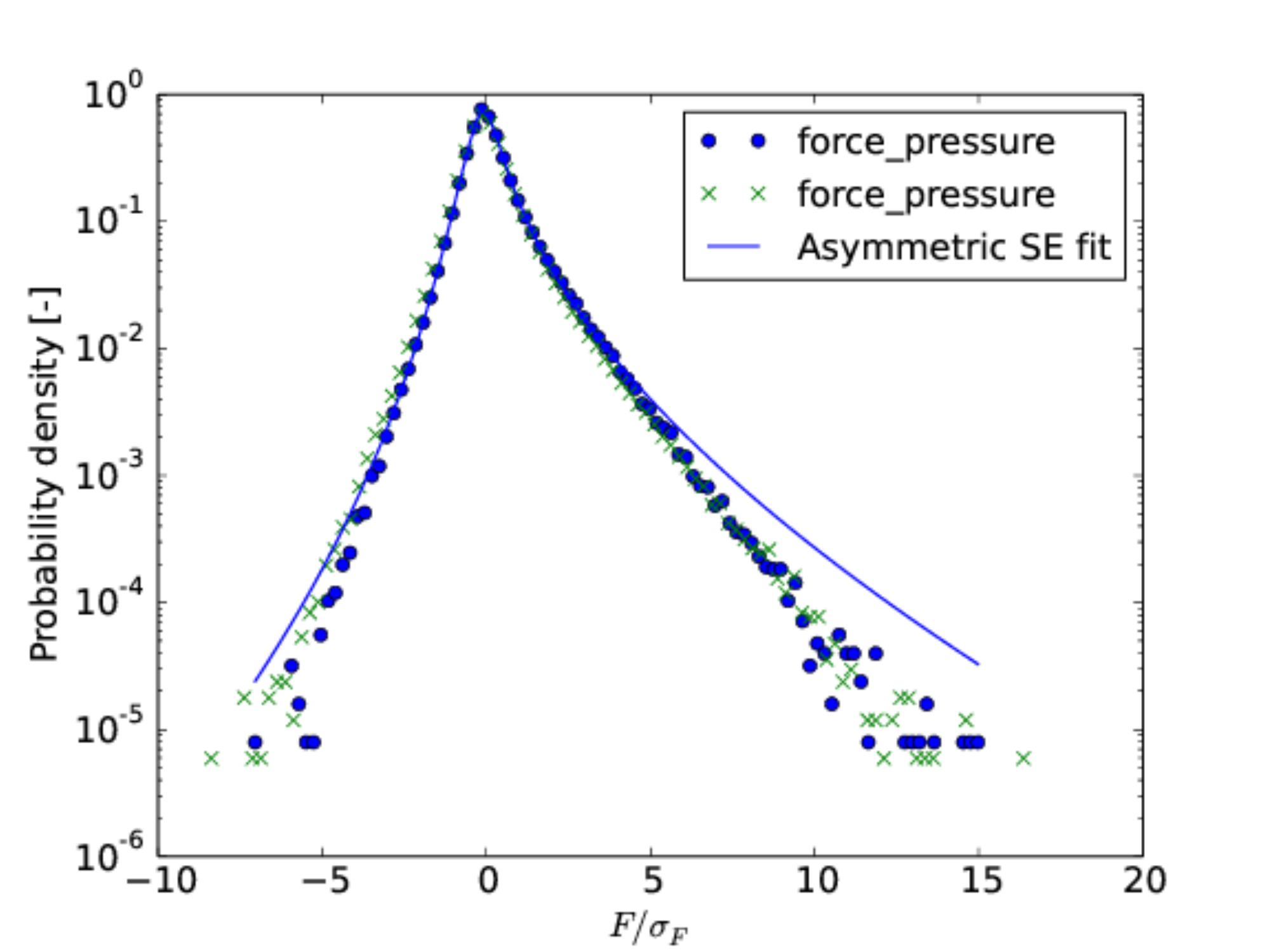}

}

\caption{Independent force terms, standardized\label{fig:Independent-force-terms}}

\end{figure}

First we consider in figure \ref{fig:Independent-force-terms} the
PDFs of the inertia force, $F_{\mathrm{I}}$ (fig. \ref{fig:Probability-distributioninertia})
and the pressure gradient force, $F_{\mathrm{p}}^{\prime}$ (fig.
\ref{fig:Normalized-probability-distribut-pressure}) for tracers
and inertial particles.

The standardized distribution of inertia force $F_{\mathrm{I}}$,
which here stands for the particle acceleration, is similar for both
tracers and inertial particles, independent of their size and density,
as reported previously \cite{calzav_accel_09}. This result is counter-intuitive,
because tracers and inertial particles are exposed to different accelerations.
Recalling that the left hand side of the equation of motion balances
the right hand side which is a sum of other forces, we emphasize that
the observed similarity of the two distributions does not come from
the same source for the inertial particles and tracers. Each type
of particles experiences different interaction forces both by amplitude
and by the combination of different terms. The similarity of the standardized
PDFs implies that the particle-fluid interaction forces are in some
sense responsible to coordinate the particle motion in the flow, keeping
the particle from developing a high relative velocity.

The standardized distributions of the pressure force term of the inertial
particles and the tracers also behave similarly. In this case, however,
the result is expected. This is because for both data sets, the distribution
is obtained by interpolation from the tracers series, and only scaled
by the particle mass, which is not seen in the standardized plot.
The only possible difference between tracers and inertial particles
in generating the pressure force PDFs is that the points of interpolation
are different, gathered from the different paths taken by tracers
and inertial particles. However, both sets of interpolation points
are large enough to map the entire flow, so that the standardized
PDFs show self-similarity.

\begin{figure}
\subfloat[Standardized interaction force ($F_{\mathrm{r}}$ in eq. \ref{eq:final})\label{fig:Standardized-interaction-force}]{\noindent \begin{centering}
\includegraphics[width=0.5\columnwidth]{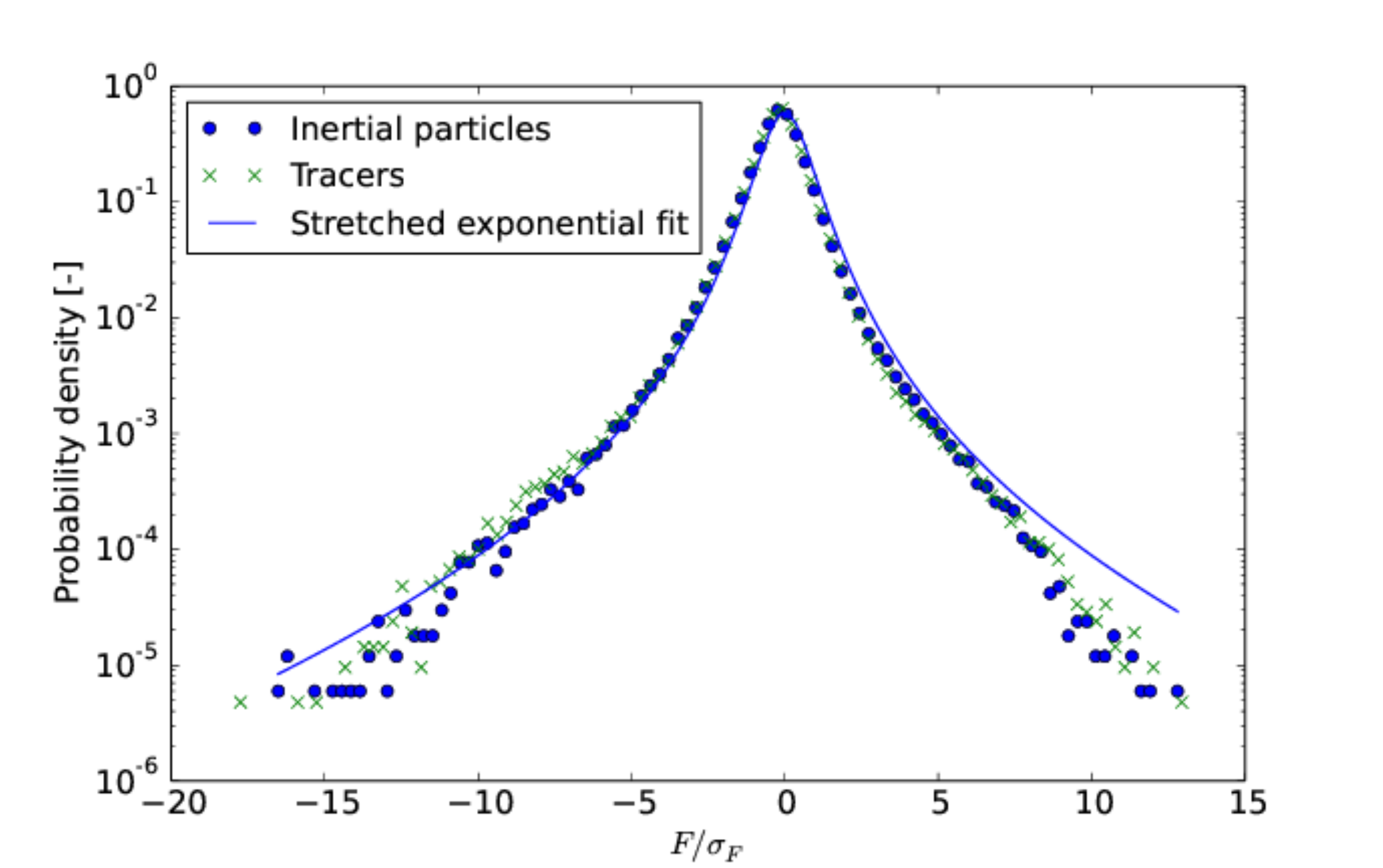}
\par\end{centering}

}

\subfloat[Drag force ($F_{\mathrm{D}}$) PDFs\label{fig:Probability-distribution-drag}]{\includegraphics[width=0.49\columnwidth]{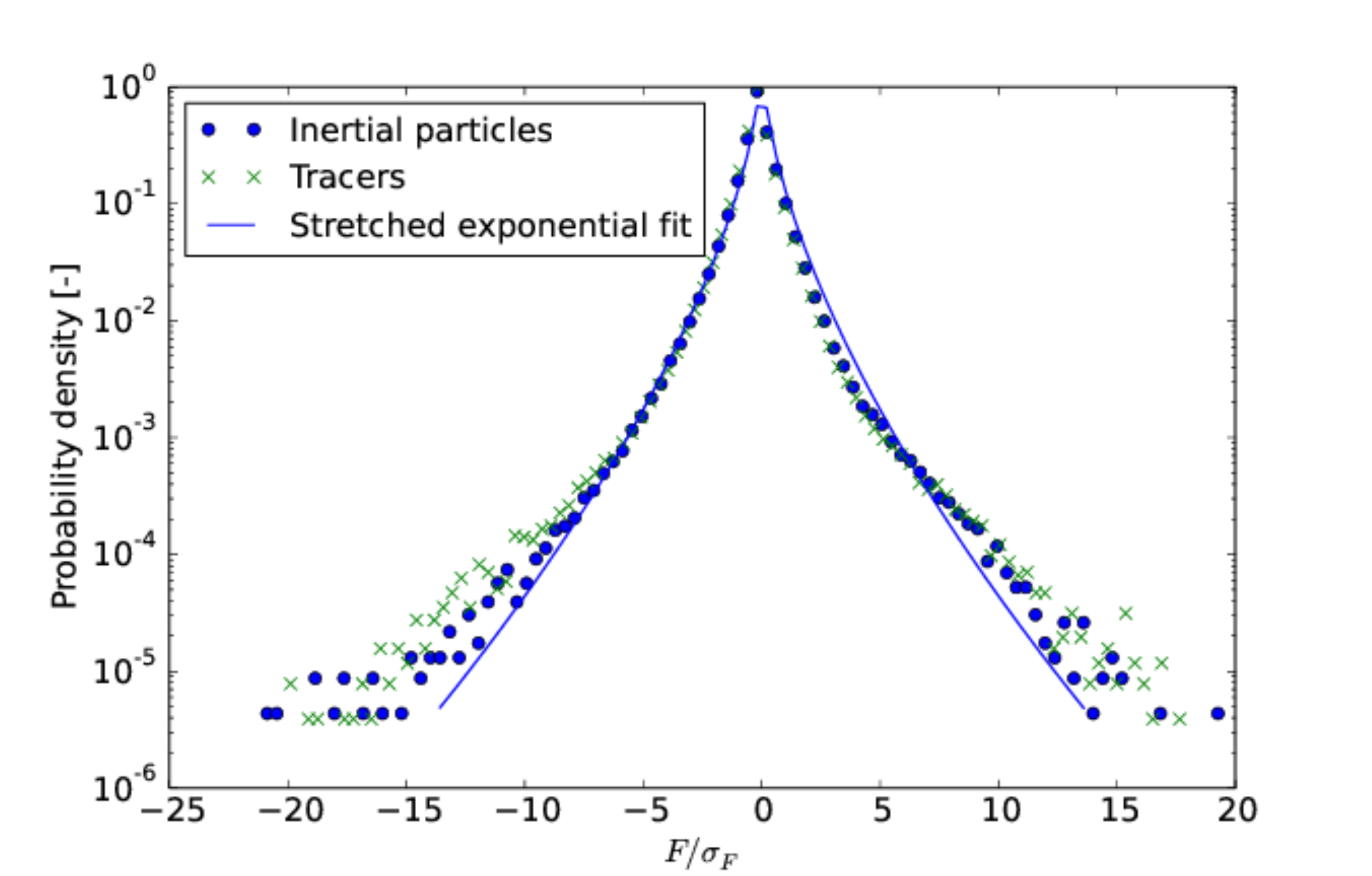}

} \subfloat[Lift force ($F_{\mathrm{L}}$) PDFs\label{fig:Normalized-probability-distribut-lift}]{\includegraphics[width=0.49\columnwidth]{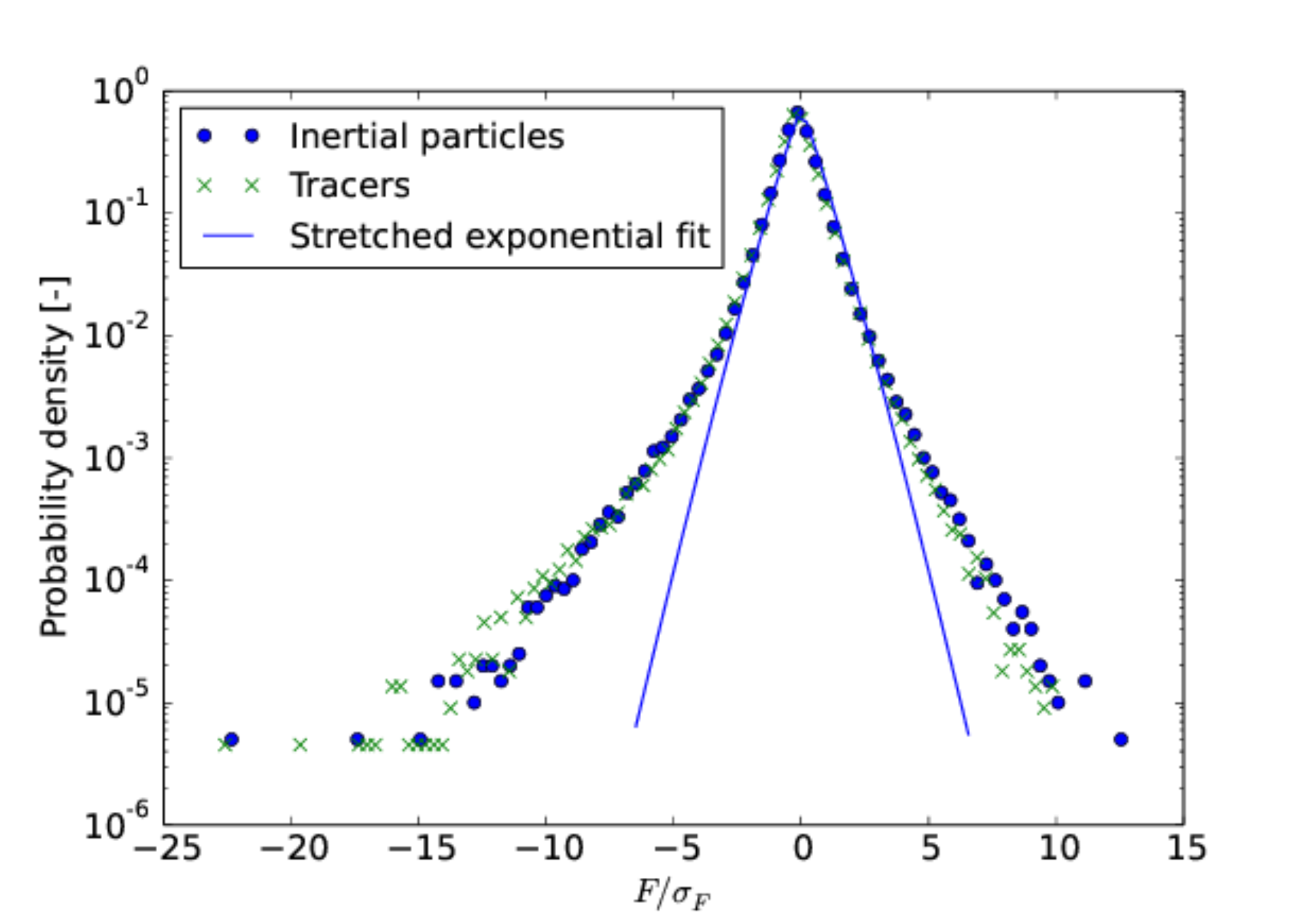}

}

\caption{Dependent force terms, standardized\label{fig:dependent-force-terms}}
\end{figure}

The similarity of both inertia (particle acceleration) and pressure
(fluid acceleration) force terms provides another peculiar result.
Despite the fact that the order of magnitude of forces are different,
and despite the different combination of particle-interaction forces
(more lift for large particles, different drag components, etc.) because
both the pressure force and the inertia force show similar PDFs, also
the the drag and lift force terms, have to show similarity in the
distributions. As we rigorously prove in appendix \ref{sec:Proof},
the self-similarity of the \emph{total} interaction force in these
conditions is a mathematical necessity if additionally the interaction
forces are statistically independent of the undisturbed fluid velocity.
This can happen if the interaction force is a function of not only
the relative velocity, but also its gradient, and possibly other factors.
Our data shows that the independence of pressure gradient force and
total interaction forces is real, as shown by the self-similarity
of the total interaction force PDF, figure \ref{fig:Standardized-interaction-force}.

\begin{figure}
\subfloat[Inertial particles]{\includegraphics[width=0.49\columnwidth]{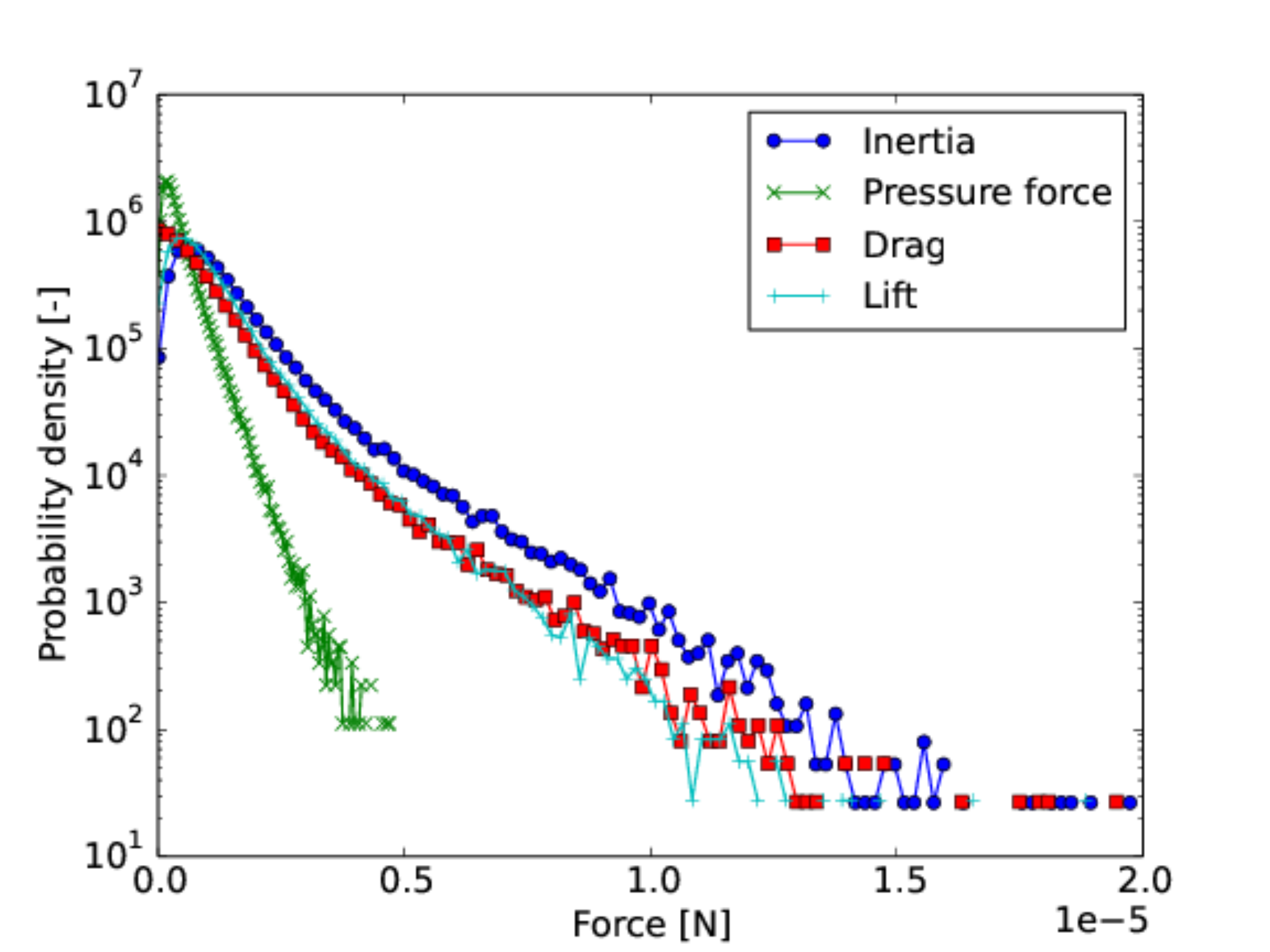}

} \subfloat[Tracers]{\includegraphics[width=0.49\columnwidth]{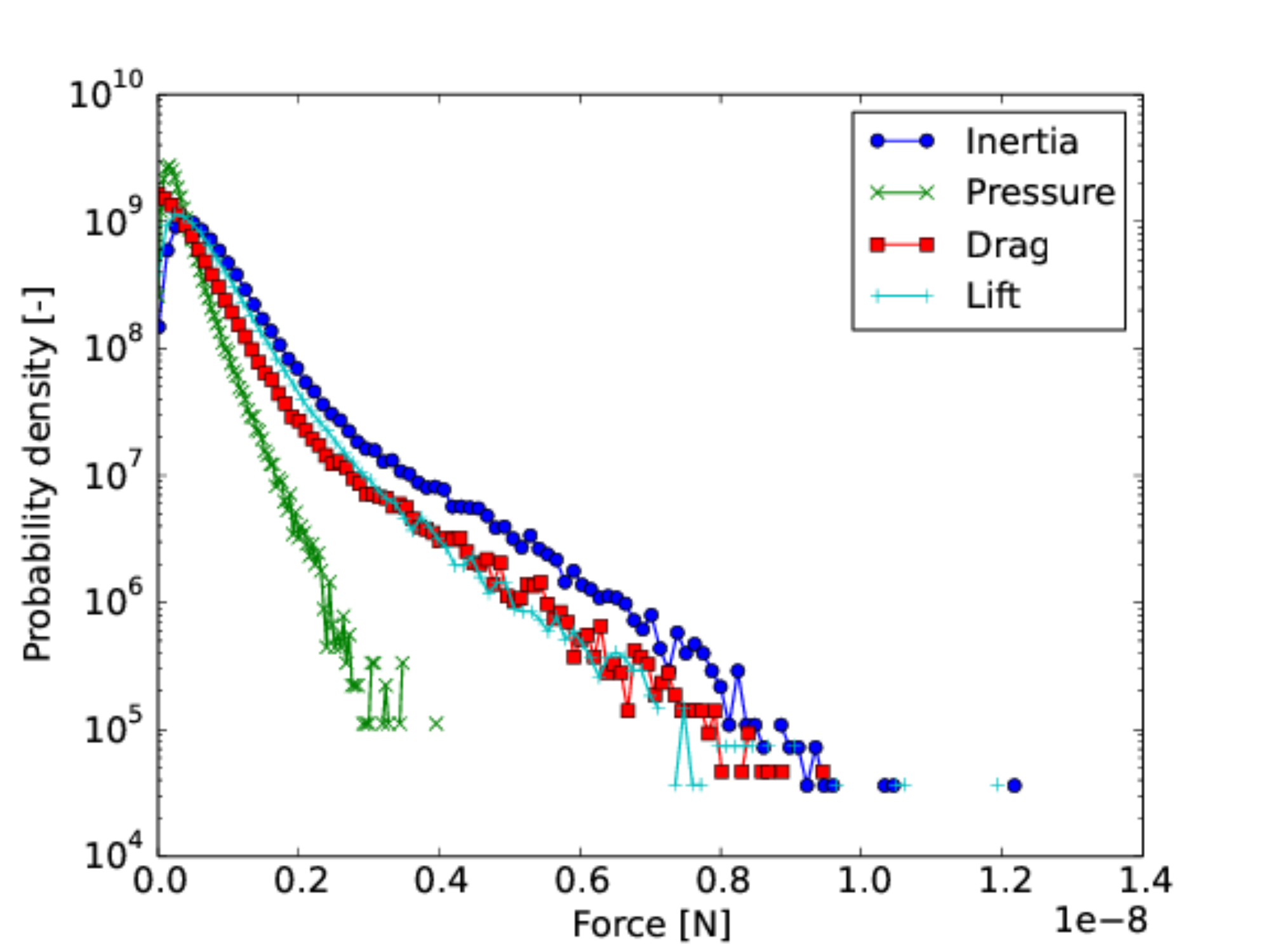}

}

\caption{Relative importance of pressure gradient and particle-fluid interaction
forces.\label{fig:Relative-importance}}

\end{figure}

What we find more surprising is that the similarity of standardized
distributions is apparent not only in the sum of interaction forces,
but also in the orthogonal components of drag (parallel to relative
velocity), fig. \ref{fig:Probability-distribution-drag}, and lift
(perpendicular to it), fig. \ref{fig:Normalized-probability-distribut-lift}.
This is a combined result of the interaction force PDFs similarity
and isotropy of the flow and low buoyancy, such that the relative
velocity and its gradient have no preferred direction. From figure
\ref{fig:Relative-importance} it becomes clear that the drag and
lift components have the same magnitude, contrary to the common assumption
that lift is small compared to drag forces. 

All of the force distributions fit a stretched exponential form \cite{voth_variance},
except for the pressure force distribution, which is skewed, and each
side fits separately to a stretched exponential with different parameters.
In isotropic turbulence, the pressure force distribution should be
symmetric, as shown by analytic considerations \cite{hs_pressure_skewed_93}
as well as simulations \cite{vy_pressure_sim}, therefore the pressure
force distribution indicates that for this experiment a certain anisotropy
is created by a secondary flow, perhaps related to the system's geometry.
In contrast, the inertia force PDFs are symmetric, as would be expected
from particles in isotropic turbulence. The anisotropy of the pressure
force is masked by the other forces acting on the particle. These
seem to be affected more by the small scales of turbulence, which
are independent of the secondary flow. Such result lends support for
the mechanism of lift suggested by Kim and Balachander \cite{kb_fluctuating_lift},
which focus on fluctuations of a size scale smaller than the particle
as the main mechanism responsible for generation of interaction forces.

Analyzing the distributions of the different force components as defined
in Eq. \ref{eq:final} we can provide here for the first time the
plot that emphasizes the relative contributions of each one of the
terms. The relative importance of the pressure gradient (fluid-driven)
versus the components of fluid-particle interaction forces (relative
velocity dependent) is given in figure \ref{fig:Relative-importance},
where not normalized distributions for both the inertial particles
and the tracers are shown. In both cases, the distribution of the
interaction forces is wider than that of the pressure force and closer
to the inertial force. The ability of the particle-fluid interaction
forces to mask out the skewness of the pressure gradient force is
a result of their relative strength.

\begin{figure}
\includegraphics[width=0.5\textwidth]{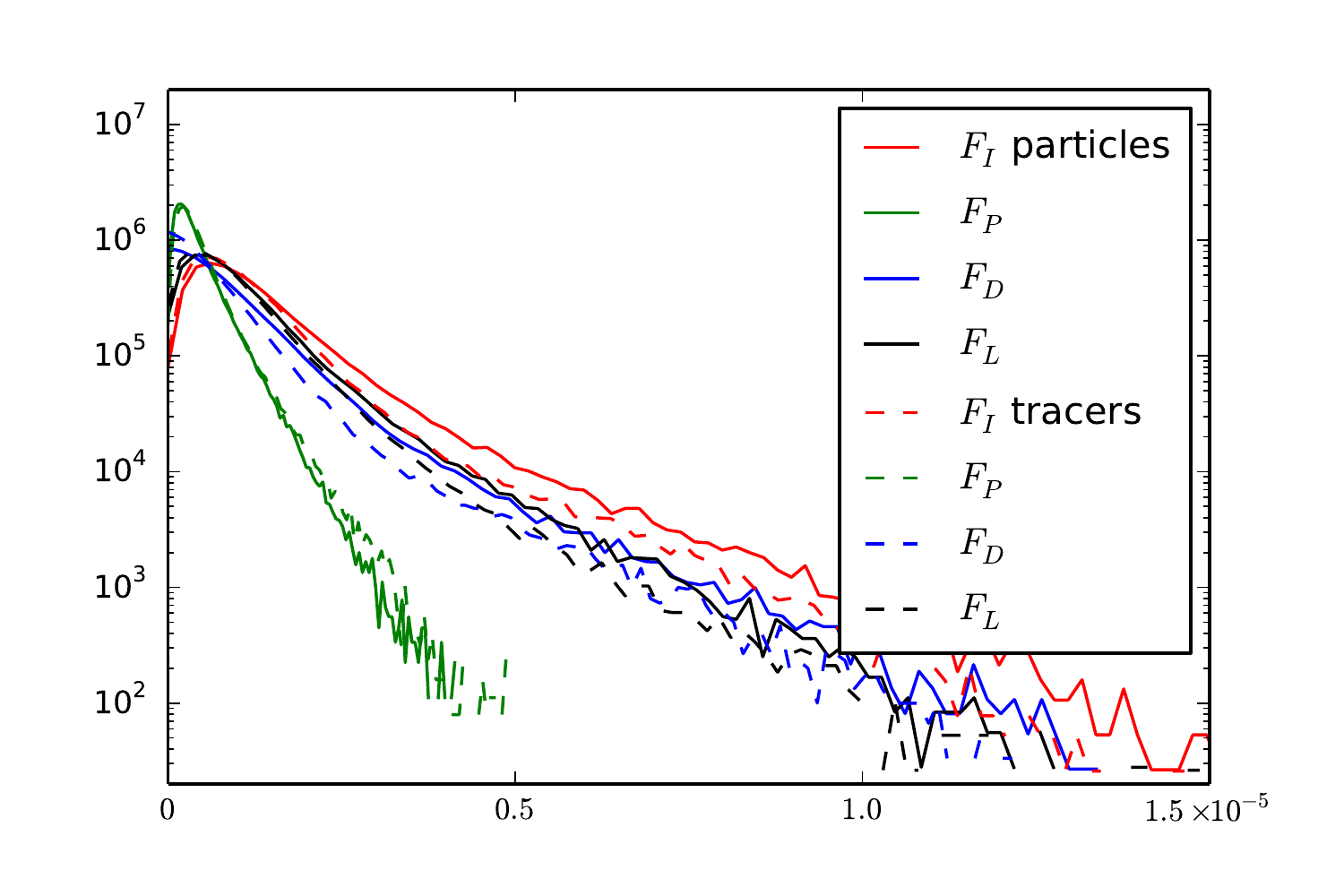}\includegraphics[width=0.5\textwidth]{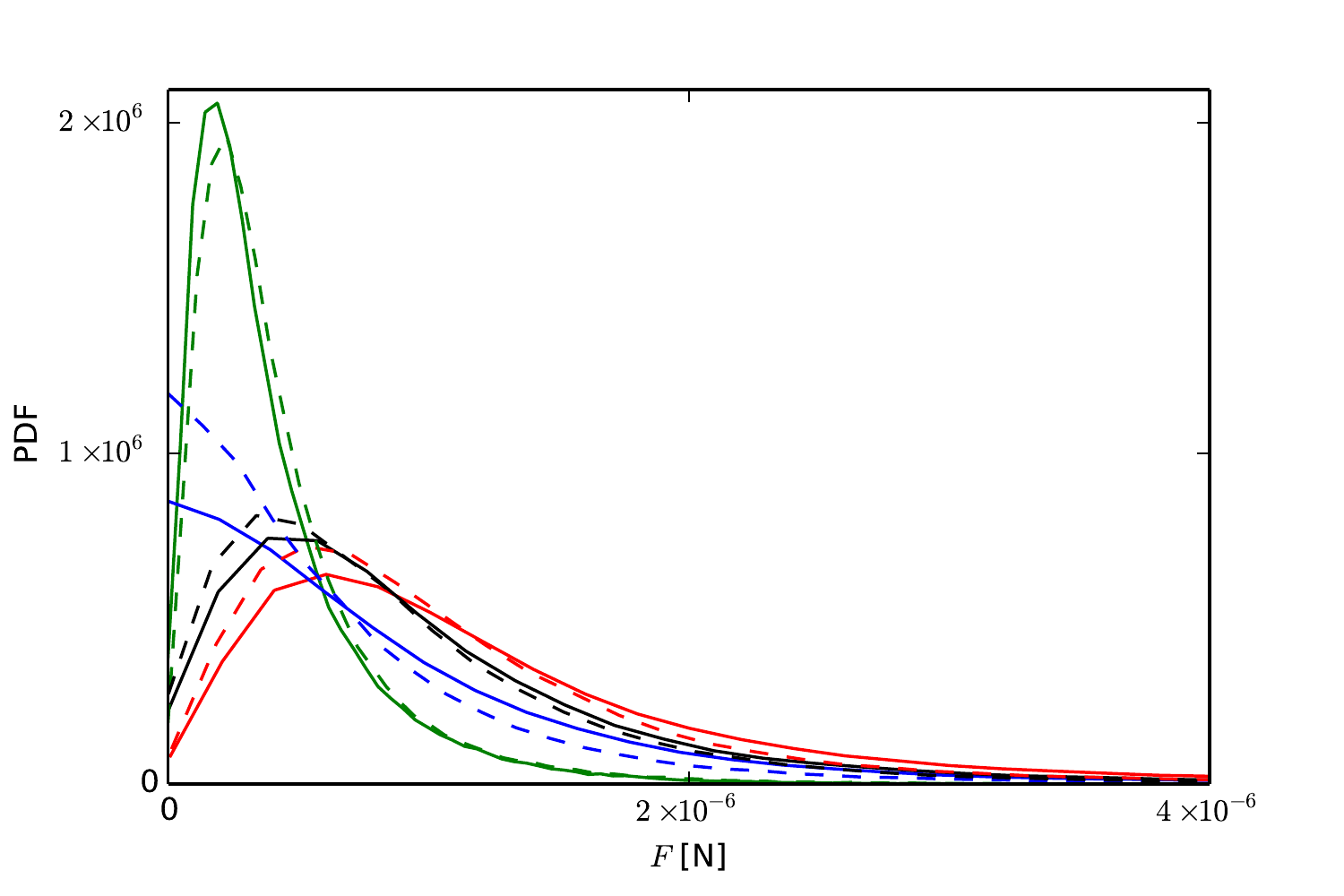}

\caption{PDFs of the forces for particles (solid curves) and tracers (dashed
curves) in logarithmic and linear scales. \label{fig:PDFs-norms}}
\end{figure}

In figure \ref{fig:PDFs-norms} we present the distributions of the
inertial particles in real values and distributions of tracers scaled
by $0.14\times10^{4}$ which is the mass ratio of the inertial particles
and tracers. The collapse shows that the distributions are very similar
in both cases, despite the three orders of magnitude differences with
few exceptions: the drag and lift that tracers experience are shifted
more towards the lower values, as compared to the inertial particles.
However, in general, we found the similarities between the distributions
to be the central (and unexpected) result of this study. 

\section{Conclusions\label{sec:Conclusions}}

The similarity of acceleration PDFs of particles with different size
and density is a relatively recent observation that helps to understand
 better  particle motion in turbulent flows. It shows that, at least
for  particles less than few times the Kolmogorov scales of the flow,
the behavior of particles can be estimated using some properties of
the turbulent flow, regardless of the particle size. Before this study
the similarity was observed only for the distributions of accelerations
of the particles.

We analysed a unique data set of directly and simultaneously measured
Lagrangian trajectories of flow tracers along with the (almost) neutrally
buoyant inertial particles which are ten times larger than the tracers
and of the order of magnitude of the Kolmogorov length scale \cite{guala_rot_disks}.
The particles and the flow were measured in a quasi-isotropic, quasi-homogeneous
turbulent flow at $Re_{\lambda}\sim250$ between counter-rotating
baffled disks. The analysis allowed a deeper look into the distributions
of the forces using a suitable formulation of particles' equation
of motion. The central point is our ability to accurately estimate
the undisturbed fluid velocity at the position of each inertial particle
and along its Lagrangian trajectory, obtained by interpolation from
neighbor tracers. After examining several interpolation methods, we
chose Inverse Distance Weighting with a parameter that minimizes the
error due to interpolation by testing the tracer ``relative velocity''.

The distributions of forces acting on inertial particles and flow
tracers in this turbulent flow indicate that both the pressure gradient
force and the particle-fluid interaction forces play a significant
role, while the interaction forces exhibit larger values. This means
that in a flow of this kind, an accurate model of the interaction
forces would be essential in order to faithfully predict particle
movement. Furthermore, the role of lift is found to be of the same
magnitude as that of drag. Thus the models of particles in turbulence,
that are not much smaller than the Kolmogorov scale, must include
a lift term.

Standardized PDFs of acceleration, expressed here as inertia force,
$F_{\mathrm{I}}$, collapse onto a single curve for both the flow
tracers and the inertial particles. The unexpected result is the similar
collapse of the distributions of the interaction forces. We derive
a mathematical expressions of the PDFs demonstrating this as a consequence
of the orthogonal decomposition and the fact that the sum is self-similar. 

The relative velocity distribution of inertial particles is broader
than that of smaller particles, but much closer to a Gaussian process.
This distribution also helps to understand  the similarity of the
drag force PDFs. We can show using our directly measured values that
the relative velocity is well correlated with the particle velocity,
emphasizing the known effect of inertia: as the flow gets faster,
the slow particle response leads to high relative velocities. The
correlation is, nevertheless, not perfect and there are also significant
relative velocity values associated with other regions of the flow.

The particle acceleration PDFs are symmetric, in contrast to the asymmetric
pressure gradient distribution. This indicates that the mechanism
generating particle-fluid interaction forces depends more on the small-scale
turbulent fluctuations, as has been recently proposed \cite{kb_fluctuating_lift}.

The present findings shall be taken with caution - one cannot expect
that the PDFs of the force terms will keep its shape similar for any
arbitrary large particle size, even for neutrally buoyant particles
such as those studied in this work. We note that although similar,
as size grows, the relative velocity distribution widens and particles
effectively filter out strong turbulent events (in terms of fluid
velocity). We believe that the self-similarity of the PDF shape is
maintained as long as the relative velocity growth can compensate
for the increased size.

\section*{Acknowledgements}

The experiments were performed at the Institute of Environmental Engineering,
ETH Zurich by Michele Guala, Klaus Hoyer and Alex Liberzon, Ref. \cite{guala_rot_disks}.
Raw data analysis was repeated by YM using the OpenPTV - an open source
particle tracking velocimetry software package, www.openpiv.net

\appendix

\section{Proof of self-similarity propagation\label{sec:Proof}}

In this section we provide a proof to the claim above, that if the
inertia force curves of two different particles in the same statistical
flow have the same standardized shape, then so will the standardized
PDF of the total particle-fluid interaction force. 

The following definitions are used for a continuous random variable
$X$: 
\begin{itemize}
\item Its standard distribution is $\sigma_{X}$.
\item The standard-deviation normalized variable is $X^{\sigma}\equiv X/\sigma_{X}$.
\item Its probability density function is $P_{X}\left(x\right)$ where $x$
is any value that can be taken by $X$.
\end{itemize}
Using this terminology, the formal theorem is thus: Let $G,D_{1},D_{2}$
be continuous independent random variables, and $m_{1},m_{2}$ scalars.
Let 
\begin{eqnarray}
I_{1} & = & m_{1}G+D_{1}\label{eq:inertia_sum}\\
I_{2} & = & m_{2}G+D_{2}
\end{eqnarray}

Theorem: If $P_{I_{1}^{\sigma}}\left(x\right)=P_{I_{2}^{\sigma}}\left(x\right)$,
then $P_{D_{1}^{\sigma}}\left(x\right)=P_{D_{2}^{\sigma}}\left(x\right)$
and the distributions hold the relation 
\[
\frac{m_{1}\sigma_{D_{1}}}{\sigma_{I_{1}}^{2}}=\frac{m_{2}\sigma_{D_{2}}}{\sigma_{I_{2}}^{2}}
\]
The specific fluid-dynamic meaning of each term is explained in section
\ref{sub:Application-to-particles}.

\subsection{Proof}

Divide eq. \ref{eq:inertia_sum} by $\sigma_{I_{1}}$ and multiply
each term $X$ on the right hand by $\frac{\sigma_{X}}{\sigma_{X}}$to
get 
\begin{equation}
I_{1}^{\sigma}=m_{1}\frac{\sigma_{G}}{\sigma_{I_{1}}}G^{\sigma}+\frac{\sigma_{D_{1}}}{\sigma_{I_{1}}}D_{1}^{\sigma}
\end{equation}

The PDF of a sum of independent random variables is the convolution
of the PDFs of the summed terms, hence 
\begin{equation}
P_{I_{1}^{\sigma}}\left(x\right)=m_{1}\frac{\sigma_{G}\sigma_{D_{1}}}{\sigma_{I_{1}}^{2}}\left(P_{G^{\sigma}}*P_{D_{1}^{\sigma}}\right)\left(x\right)\label{eq:prob_dist_sum}
\end{equation}

Apply the same procedure, mutatis mutandis, to $I_{2}$. Now substitute
into the given $P_{I_{1}^{\sigma}}\left(x\right)=P_{I_{2}^{\sigma}}\left(x\right)$
to get 
\begin{equation}
m_{1}\frac{\sigma_{G}\sigma_{D_{1}}}{\sigma_{I_{1}}^{2}}\left(P_{G^{\sigma}}*P_{D_{1}^{\sigma}}\right)\left(x\right)=m_{2}\frac{\sigma_{G}\sigma_{D_{2}}}{\sigma_{I_{2}}^{2}}\left(P_{G^{\sigma}}*P_{D_{2}^{\sigma}}\right)\left(x\right)
\end{equation}

To remove the common distribution $G$ we take the Fourier transform
of both sides, 
\begin{equation}
m_{1}\frac{\sigma_{G}\sigma_{D_{1}}}{\sigma_{I_{1}}^{2}}\tilde{P}_{G^{\sigma}}\left(\omega\right)\tilde{P}_{D_{1}^{\sigma}}\left(\omega\right)=m_{2}\frac{\sigma_{G}\sigma_{D_{2}}}{\sigma_{I_{2}}^{2}}\tilde{P}_{G^{\sigma}}\left(\omega\right)\tilde{P}_{D_{2}^{\sigma}}\left(\omega\right)
\end{equation}
so that $P_{G^{\sigma}}$ cancels out, then after the inverse transform
is applied, 
\begin{equation}
m_{1}\frac{\sigma_{G}\sigma_{D_{1}}}{\sigma_{I_{1}}^{2}}P_{D_{1}^{\sigma}}\left(x\right)=m_{2}\frac{\sigma_{G}\sigma_{D_{2}}}{\sigma_{I_{2}}^{2}}P_{D_{2}^{\sigma}}\left(x\right)
\end{equation}

Rearranging, 
\begin{equation}
\frac{P_{D_{1}^{\sigma}}\left(x\right)}{P_{D_{2}^{\sigma}}\left(x\right)}=\frac{m_{2}}{m_{1}}\frac{\sigma_{D_{2}}}{\sigma_{D_{1}}}\left(\frac{\sigma_{I_{1}}}{\sigma_{I_{2}}}\right)^{2}\equiv\alpha\label{eq:dist_ratio}
\end{equation}
where $\alpha$ is a constant.

Now, each probability density function must integrate to 1 at infinity,
\begin{eqnarray}
\int_{-\infty}^{\infty}P_{D_{1}^{\sigma}}\left(x\right)dx & = & 1\label{eq:pdf_integ}\\
\int_{-\infty}^{\infty}P_{D_{2}^{\sigma}}\left(x\right)dx & = & 1
\end{eqnarray}
and from eq. \ref{eq:dist_ratio} 
\begin{equation}
\int_{-\infty}^{\infty}P_{D_{1}^{\sigma}}\left(x\right)dx=\int_{-\infty}^{\infty}\alpha P_{D_{2}^{\sigma}}\left(x\right)dx=\alpha\int_{-\infty}^{\infty}P_{D_{2}^{\sigma}}\left(x\right)dx=\alpha
\end{equation}
So that with eq. \ref{eq:pdf_integ} we get $\alpha=1$. This allows
us to rearrange eq. \ref{eq:dist_ratio} in two ways:
\begin{itemize}
\item The lefthand side expands to $P_{D_{1}^{\sigma}}\left(x\right)=P_{D_{2}^{\sigma}}\left(x\right)$,
\item the righthand side expands to 
\[
\frac{m_{1}\sigma_{D_{1}}}{\sigma_{I_{1}}^{2}}=\frac{m_{2}\sigma_{D_{2}}}{\sigma_{I_{2}}^{2}}
\]

\end{itemize}
QED.

As an aside we note that the converse claim, $P_{D_{1}^{\sigma}}\left(x\right)=P_{D_{2}^{\sigma}}\left(x\right)\rightarrow P_{I_{1}^{\sigma}}\left(x\right)=P_{I_{2}^{\sigma}}\left(x\right)$,
is also trivially true, as immediately seen from eq. \ref{eq:prob_dist_sum}.

\subsection{Application of the proof to distributions of particles in turbulent
flow\label{sub:Application-to-particles}}

The standardized inertia force is the sum of several force terms acting
on the particle. In its simplest form, the particle's equation of
motion may be written as 
\begin{equation}
m_{\mathrm{p}}\frac{dV}{dt}=m_{F}\frac{dU}{dt}+D\label{eq:of_motion}
\end{equation}
where $D$ is the sum of forces emanating from the interaction of
the particle with the fluid, collecting surface forces such as Stokes
form drag, Saffman lift, Basset history term etc. 

In terms of the theorem above, we define 
\begin{eqnarray*}
m_{\mathrm{p}}\frac{dV}{dt} & \equiv & I\\
\frac{dU}{dt} & \equiv & G
\end{eqnarray*}

When the same flow contains two types of particles, e.g. small, neutrally
buoyant tracers together with larger inertial particles, we have $I_{1,2},m_{F1,2},D_{1,2}$.
But $G$ is common to both particle sizes, as they both share the
same flow. Substituting these terms into eq. \ref{eq:of_motion} we
get eqns. \ref{eq:inertia_sum}.

The observation that acceleration PDFs are invariant translates into
$P_{I_{1}^{\sigma}}\left(x\right)=P_{I_{2}^{\sigma}}\left(x\right)$,
which is the condition for the propagation of invariance theorem,
yielding $P_{D_{1}^{\sigma}}\left(x\right)=P_{D_{2}^{\sigma}}\left(x\right)$.

The conclusion is that the PDF of total particle-fluid interaction
forces must show standardized invariance, independent of the pressure
gradient force, if standardized acceleration PDFs are invariant.

\bibliographystyle{unsrt}
\bibliography{refs}

\end{document}